# LLM Agents as Static Level-k Players in Behavioural Games

## Design choices, distributional fidelity, and the reproduction of human strategic play

Po Han Teo

Independent Researcher, Singapore

aaronteopohan@hotmail.com

Google Scholar: https://scholar.google.com/citations?user=UtMTbCYAAAAJ

**Abstract**

Large Language Models (LLMs) are increasingly used as stand-ins in behavioural games. These stand-ins rely on the assumption that the LLM's distribution of choices meaningfully matches how humans play the same game. This study tests that assumption through two games. The first is a p-beauty contest, and the second one is a public goods game. The study first investigates five local-model settings within the same model family. These settings are varied together in a 360-cell factorial, which balances temperature, scale (0.5–32B), quantisation, instruct vs base, and framing. Each cell's distribution is then compared against whole choice distributions in published human data. Each deployment setting, except for quantisation, governs a different aspect of fidelity. Mechanically, while the dispersion of human players can be somewhat recovered through deployment settings, the strategic process behind it cannot. Through the lens of the level-k cognitive theory, we find that LLMs act as static, category-retrieved level-k players, where k is set by the model scale. The models also do not run within-game belief-updating or backward induction throughout multiple-round horizon settings. While human contributions decayed in the public goods game, LLMs stayed flat or rose at every scale. When the horizon test was administered, LLMs were more cooperative under an indefinite horizon compared to a finite one. However, LLMs ignore their relative round position, so no last-round defection was displayed. This implies that LLMs retrieved levels relative to the horizon category rather than working out iteratively from the specific game setting.

Researchers are increasingly using LLMs as stand-in experimental subjects. Compared to human recruitment, LLMs as stand-in are cheaper, faster, and have fewer operational constraints [Argyle, 2023; Horton, 2023; Aher, 2023]. When provided with a task and an identity, a single model is able to reproduce a distribution of choices that can be read as behavioural predictions [Mei et al., 2024]. However, these approaches rest on the assumption that the model's distribution is faithful to human behaviour in the same setting. This assumption is demanding as human distributions are a result of structures and strategies. An observed distribution of human behaviour is produced by a spread of reasoning depths, and is even more pertinent when it comes to behavioural game-theoretic settings. As such, a faithful synthetic subject has to reproduce both the underlying structure and the human distribution.

This human distribution is a reflection of both Nash strategic reasoning and bounded rationality that underlies behavioural economics. However, recent studies find that LLM capability works against human fidelity, where more capable LLMs reason towards classical Nash equilibrium, while deviating from actual human behaviour. Therefore, there is a need to truly understand if there are design choices that can induce certain behavioural changes, and if these behaviours go through cognitive and strategic reasoning mechanisms.

One hypothesis to reconcile the difference in behaviours between LLMs and humans is that LLMs are static, category-retrieved level-k players. The intuitive mechanism for a LLM would be that its reasoning level, set during training, is tied to a game’s category. Thereafter, the retrieval remains static, and the LLMs hold it across multiple rounds. This retrieval is not belief-updated between rounds, nor is there actual backward induction that humans might do. To test this hypothesis, both the mechanical deployment settings of the LLM and the cognitive strategic reasoning has to be analysed.

To make the analyses more robust, we chose two games. The first is the one-shot p-beauty contest to test reasoning depth. The second is a repeated public goods game with punishment mechanics to test social dilemma and cooperation. The first round of the public goods game can also serve as a cold-start anchor to corroborate with the one-shot game. The deployment settings are then varied in a balanced factorial over a single model family (Qwen 2.5), with fidelity measurements done at the distributional level. Through the mechanical deployment study, we can contribute to the foundational understanding on how synthetic subjects alter their behaviour according to mechanical settings. After the mechanical study, we will then look at how strategic reasoning occurs in multiple round games by altering the horizon settings.

# Related Literature

Existing literature on LLMs as game-theoretic experimental subjects can be categorised into two main themes. The first theme is that LLMs, on aggregate, do reproduce human-like behavioural distributions. The second theme, on the other hand, shows that LLMs become more deterministic when model capability increases. Within the broad themes, there are some differences in the literature when it comes to the nuance of human-faithfulness in simulation.

## Theme 1: LLMs look human in aggregate

Literature in this theme has shown that with demographic conditioning, LLMs can reproduce human subgroup response distributions [Argyle, 2023]. Classical game-theoretic effects can also be replicated when a model is provided with endowments and preferences. These include the Charness and Rabin social preferences [Charness & Rabin, 2002], the endowment effect [Kahneman et al., 1991], and the homo-silicus cases [Horton, 2023]. With non-local models, a model such as ChatGPT-4 reproduces human distributions across trust, fairness, and cooperation games [Mei et al., 2024]. However, these simulations are prompt and sample-sensitive [Aher, 2023]. At the same time, deviations in simulations suffer from directional skews, with some studies showing more cooperative and altruistic simulations when compared to human ground truth [Mei et al., 2024].

## Theme 2: capability yields deterministic play

Literature in this theme has shown that the more capable an LLM is, the more deterministic the results become. Distributions start to disappear as LLMs hone in on undeviated strategies. For example, a capable model does well in self-interest games such as the Iterated Prisoner's Dilemma, but fails in coordination games (Battle of the Sexes) [Akata et al., 2025]. Reasoning-tuned models also free-ride more in public goods [Li & Shirado, 2025; Guzman Piedrahita et al., 2025], implying a return to classical inefficient Nash equilibria. In Affonso's [2026] study, models converge on competitive rationality, and diverge on cooperation. Literature in this theme points at an intuition where model capability pulls play toward equilibrium, away from human deviation.

Despite the two themes, there remain gaps when it comes to model fidelity. Models are susceptible to contextual prompting differences, even after accounting for model scale [Jia et al., 2025]. Existing literature on model benchmark tests is also breadth-wide, with the largest benchmark spanning model vendors [Affonso, 2026]. A breadth-wide test, while offering vendor comparative variance, suffers when it comes to isolating the effects of scale. In a lot of social science settings, deployment settings of LLMs tend to be treated as an after-thought or a nuisance rather than an object of study [Aher, 2023]. Fidelity in these studies was also operationalised and scored on means vs meta-analytic rates, not whole distributions. More importantly from a social science perspective, deviations between LLM simulations and human ground truth are often catalogued [Huynh et al., 2025] but left unexplained through a behavioural lens, which means there is a lack of reconciliation between process (human) versus retrieval (LLMs).

## The closest prior work and the present design

Two studies sit closest to ours. The first is a benchmark by Affonso [2026], which runs twenty-five models across thirty-eight games and scores their cooperation against meta-analytic human rates. Cooperation varies forty-eight-fold across them. These same models also collapse to defection at the last round of a finite game. The second study takes a different angle, running humans and frontier models through beauty-contest scenarios to see how close the models land to the human anchor [Alekseenko et al., 2025]. They land far short, guessing around 18 against the human 37. Both studies point the same way: capable models play near the equilibrium, and people do not.

Compared to the aforementioned closest prior works, our study differs in four ways. Firstly, we go for a depth-wise approach by choosing a single open-weight family (Qwen 2.5), and varying the scale. This approach controls for vendor confounds on scale, training, and alignment. Secondly, we manipulate the game horizon in the prompting, which differs from past work where only a finite horizon was provided. By comparing the finite game versus an indefinite one with a continuation probability, we check if the LLM is capable of strategizing differently. Thirdly, we get the LLM to explicitly state a belief about the other players and record the best response, rather than simply reacting to observed play in previous rounds. This method makes the conditional cooperation explicit in the LLM. Lastly, we compare the distributions to human ground truth. This is done in order to see if we can fully replicate distributions rather than moments in metrics. All four ways are also varied across LLM deployment settings, namely temperature, quantisation, and base-versus-instruct. These allow us to have a more holistic mechanical understanding in addition to the game-theoretic lens.

## Theory: Level-k Reasoning and the Choices That Tune It

It is the human departure from the Nash equilibrium that makes behavioural games useful. Behavioural games are diagnostic because human play leaves equilibrium in a structured, repeated way. For example, in a beauty contest, the Nash equilibrium is zero in a single round under common knowledge of rationality, but people often make a guess above it. This deviation implies shallower depths of iterated reasoning compared to a fully extended recursion [Nagel, 1995; Camerer, Ho & Chong, 2004]. In a public goods game, the dominant strategy is to contribute nothing in the beginning, but human players tend to contribute around half their endowment in round one, decaying slowly subsequently. This implies additional fairness and conditional cooperation in human strategic thought processes [Fischbacher & Gachter, 2010]. In both examples, human behaviour is also probabilistic rather than a single deterministic point, spreading across reasoning depths or cooperative types. This is the structural regularity that a synthetic subject has to reproduce on a distributional level. Taken together across games, human play splits into a strategic component and a structured residual that is the deviation [Teo, 2026]. While LLMs can somewhat retrieve strategy, the structured residual requires investigation.

When it comes to strategic reasoning, cognitive hierarchy theory can be used as the lens to account for human distribution in behavioural games. A level-0 player is non-strategic, a level-1 best-responds to level-0, a level-2 to level-1, and so on. Nash equilibrium becomes an asymptotic infinite-depth limit. This results in a mixture of multiple levels that gives human play its spread. This spread is a structured regularity that has to be kept distinct from other residuals, such as decision noise. For the LLM to be able to reproduce human distribution, strategic type heterogeneity [Camerer, Ho & Chong, 2004] must be reproduced, rather than phenomena such as trembling-hand/quantal responses [McKelvey & Palfrey, 1995].

In multiple round games, level-k can become static or dynamic. A static level-k is a level that is chosen once throughout a game. A dynamic level-k, on the other hand, runs recursions over rounds through mechanisms such as belief updating, convergence, or backward induction. Human regularities should reside in the dynamic level-k. For an LLM, retrieval of the appropriate level-k resides on a separate axis, where the game's category is the primary determinant. This is in contrast to computing a level from the specific game in front of it. Static and dynamic level-k, as well as the retrieval-versus-computation modes, will serve as the lens to interpret subsequent analyses in this study.

The controllable settings of a local model, sampling temperature, parameter scale, quantisation, instruction tuning, and prompt framing, are the levers a researcher actually has when deploying LLM agents as subjects. Treating them as candidate control parameters of the level-k quantities above yields six hypotheses. Each names the coordinate a setting should move, and each is testable against the human benchmark.

H1 (depth). Reasoning depth rises with scale. Larger models place themselves higher on the ladder, opening lower in the beauty contest and conditioning more steeply in the public goods game, and past some scale they reason toward the equilibrium and away from the human deviation.

H2 (static play). Within a repeated game the agent holds its level. Contribution trajectories do not decline across rounds as human play does, at any scale, because the agent does not revise its belief about others as the game reveals it.

H3 (retrieval, not computation). The agent responds to a game's global category but not to its local structure. Cooperation tracks the finite-versus-indefinite horizon, a single retrievable association, yet an explicit final round triggers no backward-induction defection.

H4 (alignment). Instruction tuning raises the level a model opens at, relative to its base counterpart, and acts on the opening rather than on the trajectory.

H5 (temperature). Temperature widens the dispersion of a single specification around its one level, but cannot place mass at several levels at once, so it reproduces the magnitude of human spread without its between-type structure.

H6 (quantisation). Once scale is held fixed, quantisation has no systematic effect on any behavioural coordinate.

H2 and H3 are the central test, since they ask whether the agent runs any of the within-game dynamics that generate human play, the round-to-round revision of beliefs and the unravelling by backward induction at a known end. The remaining four map the static parameters that the design choices set.

# Experimental Design

For the mechanical analyses, the study design varies the controllable settings of a local model in a balanced factorial, and simulates the game in accordance with human-player instructions. Each factorial cell is one combination of settings, and is run by a group of synthetic agents whose pooled choices form a distribution. The factorial cell simulations are then analysed for behavioural fidelity. Fidelity is operationalised as the distance from a factorial cell distribution to a human target.

To ensure that the findings are more generalisable rather than being an artefact of a single paradigm, two games are chosen to anchor the factorial. The choices are made based on the games' opposite poles. The first game is a one-shot p-beauty contest to represent reasoning depth strategies. The second game is the repeated public goods game, which represents a social dilemma. Both games differ on the two pertinent axes for the level-k reading: the horizon (one-shot versus repeated) and the type of reasoning deviation (reasoning depth versus cooperation/competition). Where both games are compared, the comparison is taken at round 1 of the repeated public goods game, where the cold start happens before any learning or aggregating mechanisms have kicked in. This then matches the one-shot game.

The public goods game follows the voluntary contribution mechanism of Herrmann, Thoeni and Gachter [2008]. In their study, the setting is a four-player fixed group, lasting ten rounds, with twenty tokens of endowment per round. Every token in the shared public pot returns 0.4 to each of the four. The Nash dominant strategy is to contribute nothing. However, the human regularity finding in the study shows that players start to contribute near half their endowment, then decay slowly toward (but not to) full free-riding [Isaac & Walker, 1988; Fischbacher & Gachter, 2010]. This pattern is stable across many studies [Zelmer, 2003]. In Herrmann, Thoeni and Gachter's study, however, a punishment condition adds a second stage. After seeing each other's contributions, players may assign costly sanctions. This punitive mechanism increases cooperation in the human data [Fehr & Gachter, 2002]. Our simulated subjects see only their own history and the other subjects' anonymous contributions. They never observe the human data they are compared against. This is done to measure emergent rather than replicative behaviours.

The p-beauty contest follows Nagel's [1995] two-thirds variant. Players choose a number from 0–100, and the winner is the one who chooses closest to two-thirds of the group mean. The equilibrium is zero under common knowledge of rationality by iterated reasoning. However, humans do not start at zero. The original study found that first-round guesses cluster near 37, at shallow iterated-reasoning levels. The results imply that the human players' guesses reflect one or two applications of the two-thirds multiplier from a naive anchor of 50 [Bosch-Domenech et al., 2002]. Our study simulates the game as a single shot. As such, the fidelity target is the distribution of opening guesses, which is comparable to the cold-start round 1 of the public goods game.

## The factorial

A single family of models, Qwen 2.5, is chosen to control for vendor effects, such as the training corpus and the recipe for alignment. Under this control condition, the difference across the ladder of settings can then be interpreted as an effect of scale rather than differences in provenance across vendors.

| Factor | Levels |
| --- | --- |
| Model scale | 0.5B, 1.5B, 3B, 7B, 14B, 32B (single dense family) |

| Factor | Levels |
|---|---|
| Sampling temperature | 0.2, 0.5, 0.7, 1.0, 1.3 |
| Quantisation | four-bit, eight-bit |
| Alignment | base, instruction-tuned |
| Prompt framing | neutral, cooperative, competitive |

The synthetic subjects are initialised agents; each agent is one instance of a local open-weight model on the researcher's own hardware rather than a vendor API. This is done to make quantisation and sampling controllable. The agent count per cell matches the human sample of the corresponding study (24 groups of 4 in the public goods game, and 30 in the beauty contest). The matching of agent counts allows the simulated distribution to be more comparable to the human ground truth distributions in the respective original papers.

In accordance with the hypotheses above, the factorial tests the following: scale carries H1, alignment carries H4, temperature carries H5, quantisation carries H6, with framing as a contextual control. The round-by-round trajectory tests H2, where a static player has flat contributions but a dynamic player has a decaying one. Fidelity measurements are made at the distribution level rather than per individual. This is because simulated agents are stochastic but human players are heterogeneous. There is no good basis for one-to-one matching. The stochasticity in agents serves as the behavioural variance of simulated agents for distributional comparisons. Temperature is also held above zero throughout, with a near-zero setting kept only to show the distributional collapse. A zero temperature simply results in overly deterministic values that naturally fail any comparison to a dispersed human distribution. The distance from a model cell to its human target uses the two-sample Kolmogorov-Smirnov statistic, and the Wasserstein distance as measurement modes. The public goods game has round-level human data, where the St. Gallen open deposit, n = 96 per round [Herrmann et al., 2008], is scored as per-round KS and Wasserstein statistics. However, the one-shot beauty contest only has published guess distributions [Nagel, 1995; Bosch-Domenech et al., 2002], so fidelity in this paradigm is the match in dispersion, and in where the mass concentrates across level-k anchors. It is not measured as subject-level distances.

## Mechanism probes

After the mechanical study on the deployment settings, the cognitive and strategic reasoning investigation is done on a reduced ladder of deployment settings. The reduction is because the earlier study was completed before this cognitive and strategic reasoning investigation, and the reduced ladder focuses on the scales where a statistically significant effect is observed. In this study, we present an agent with a stated average contribution of the other agents, and record the agent's response to the information. This separates the agent's best-response function from its dynamic deployment [Fischbacher et al., 2001]. This allows us to test H2 in isolation from H3. H3 is instead tested in two parts. For the cooperation mechanism, three framings of the same public goods game are used. These include a known finite horizon, an indefinite horizon with a stated continuation probability, and an unframed one. An agent that is a reasoner would track the horizon, but an agent that retrieves would not. To elicit the existence of backward induction mechanics (the second part of H3), a version of the public goods game makes the last round explicit. If the agent could achieve backward induction, defection would happen, and vice versa. All the horizon manipulations are then compared against the same human ground truth as the main factorial.

# Results I: What Governs the Level-k Parameters

Beyond models, design choices determine where in the cognitive hierarchy agents retrieve from. They also determine the level at which agents start the game, as well as the depth to which agents reason. By having a complete and balanced factorial over the five design choices, the partial effect of each design choice can be elicited. We ran the factorial dataset of 360 cells through an OLS, and chose different dependent variables to identify the various behavioural coordinates. The opening level is measured as the round-1 contribution. Decline across the rounds is measured as a proportion of the round-1 contribution. The decline is measured as a proportion in order to delineate it from raw numerical slack: a higher opening contribution would naturally allow a steeper fall on a bounded scale, and raw values would confound numerical slack with actual behavioural decay. The settling level is defined as the mean contribution over rounds 6 to 10, normalised by the contribution range.

Based on the results, different design choices move different behavioural coordinates.

| Design choice | Round-1 dispersion | Opening level | Settling level | Proportional decline |
|---|---|---|---|---|
| Model scale (log params) | +0.025*** (0.005) | -0.010 (0.007) | -0.031*** (0.006) | +0.054** (0.020) |
| Sampling temperature | +0.032*** (0.005) | +0.020** (0.007) | -0.013* (0.006) | +0.208*** (0.020) |
| Quantisation (8-bit vs 4-bit) | +0.005 (0.005) | -0.004 (0.007) | -0.005 (0.006) | -0.003 (0.020) |
| Alignment (base vs instruct) | -0.001 (0.005) | -0.120*** (0.007) | -0.080*** (0.006) | -0.211*** (0.020) |
| Framing (cooperative vs neutral) | +0.009 (0.006) | +0.012 (0.008) | +0.001 (0.007) | +0.023 (0.023) |
| Framing (competitive vs neutral) | +0.011 (0.006) | +0.013 (0.008) | +0.013 (0.007) | -0.012 (0.023) |
| N | 360 | 360 | 360 | 360 |
| R-squared | 0.15 | 0.50 | 0.35 | 0.39 |
| Adjusted R-squared | 0.14 | 0.49 | 0.34 | 0.38 |
| F (6, 353) | 10.6 | 59.2 | 31.7 | 37.4 |

*Table 1. Standardised partial effects of each design choice on the behavioural coordinates of the public goods game, from a separate ordinary-least-squares regression per coordinate on the complete 360-cell balanced factorial (condition N). Each entry is the standardised coefficient with its standard error in parentheses; predictors are standardised. Significance: *** p<0.001, ** p<0.01, * p<0.05; all four model F tests are significant at p<0.001.*

When it comes to alignment, instruction-tuned models open at a higher level than base models. This is the largest effect on the opening level ($\beta = -0.120$, 95% CI [-0.133, -0.107], $p<0.001$). The possible explanatory mechanism would be that instruction tuning elicits a cooperative prior, which in turn retrieves a disposition to contribute. On the other hand, the base model (completing text without that prior) lacks this additional retrieval. Instruction-tuned models also settle a little higher than base models. At the same time, they show more proportional decline than base models ($\beta = -0.211$ for base versus instruct, $p<0.001$), their round trend running nearer the Nash direction, though like every cell it climbs rather than falls.

For parameter scales, an increase in scale lowers the settling level ($\beta = -0.031$, 95% CI [-0.043, -0.018], $p<0.001$) and increases the proportional decline ($\beta = +0.054$, $p<0.01$). Taken together, higher scale gravitates towards Nash strategic reasoning, the free-riding direction in the public goods game. At the cold start, scale does not move the opening level ($p = 0.13$), but it does widen the round-1 dispersion ($\beta = +0.025$, $p<0.001$), an effect separate from temperature's.

Raising model temperature understandably widens round-1 dispersion ($\beta = +0.032$, $p<0.001$). The dispersion effect is expected since temperature increases sampling noise in LLMs. Temperature increase also increases proportional decline ($\beta = +0.208$, 95% CI [+0.168, +0.248], $p<0.001$). On the settling levels, the effect of increasing temperature is borderline significant ($\beta = -0.013$, $p = 0.045$). Taken together, the effect of temperature on decline is separable from opening level noise. As rounds progress, the effect of temperature becomes weaker.

For quantisation, its effect is not statistically significant once size is held fixed. Across four-bit and eight-bit models, quantisation moves none of the behavioural coordinates ($|\beta| \leq 0.005$, $p > 0.3$). During simulations, smaller models collapsed onto equilibrium under four-bit compression, but this effect disappears at higher parameter scales. The collapse is thus more attributable to scale than to quantisation. Intuitively, quantisation precision is not an important parameter for the population mean if held constant, as it does not have systematic effects on the ladder of models.

For framing, there are no significant main effects. However, the competitive-framing effect reverses with scale: the competitive × scale interaction on the opening level is significant ($\beta = -0.041$, 95% CI [-0.055, -0.027], $p<0.001$), while the cooperative × scale interaction is not ($p = 0.69$). Smaller models contribute more under competitive framing than cooperative framing, while larger models go in the Nash strategic direction, contributing less under competitive framing [Liberman et al., 2004]. The interaction effect indicates correctness of a model's comprehension of framing, which happens only when models are large enough.

Read together, the four design choices serve as the control parameters of the static level-k player. Alignment sets the level it opens at (H4), scale sets the depth (H1), temperature sets the noise (H5), and quantisation, held fixed, sets nothing (H6).

While deployment settings do somewhat capture strategic reasoning across different behavioural coordinates, they do not capture the mechanisms that created the human dispersion. Across the factorial, no single deployment cell reproduces the human distribution: the Kolmogorov-Smirnov distance between a cell and the St. Gallen subjects holds between 0.40 and 0.52 at every scale, lacking a zero-distance faithful match. The dispersion in human data occurs because humans are heterogeneous. A set of identical synthetic agents creates a spike at one level-k, and raising the temperature merely widens that spike around a single mode without placing other modal mass in the distribution. As such, the resulting variance from changing deployment settings is more attributable to decision noise than to heterogeneous players with strategic deviation. There is thus still a need to make the synthetic agents heterogeneous to capture true human dispersion.

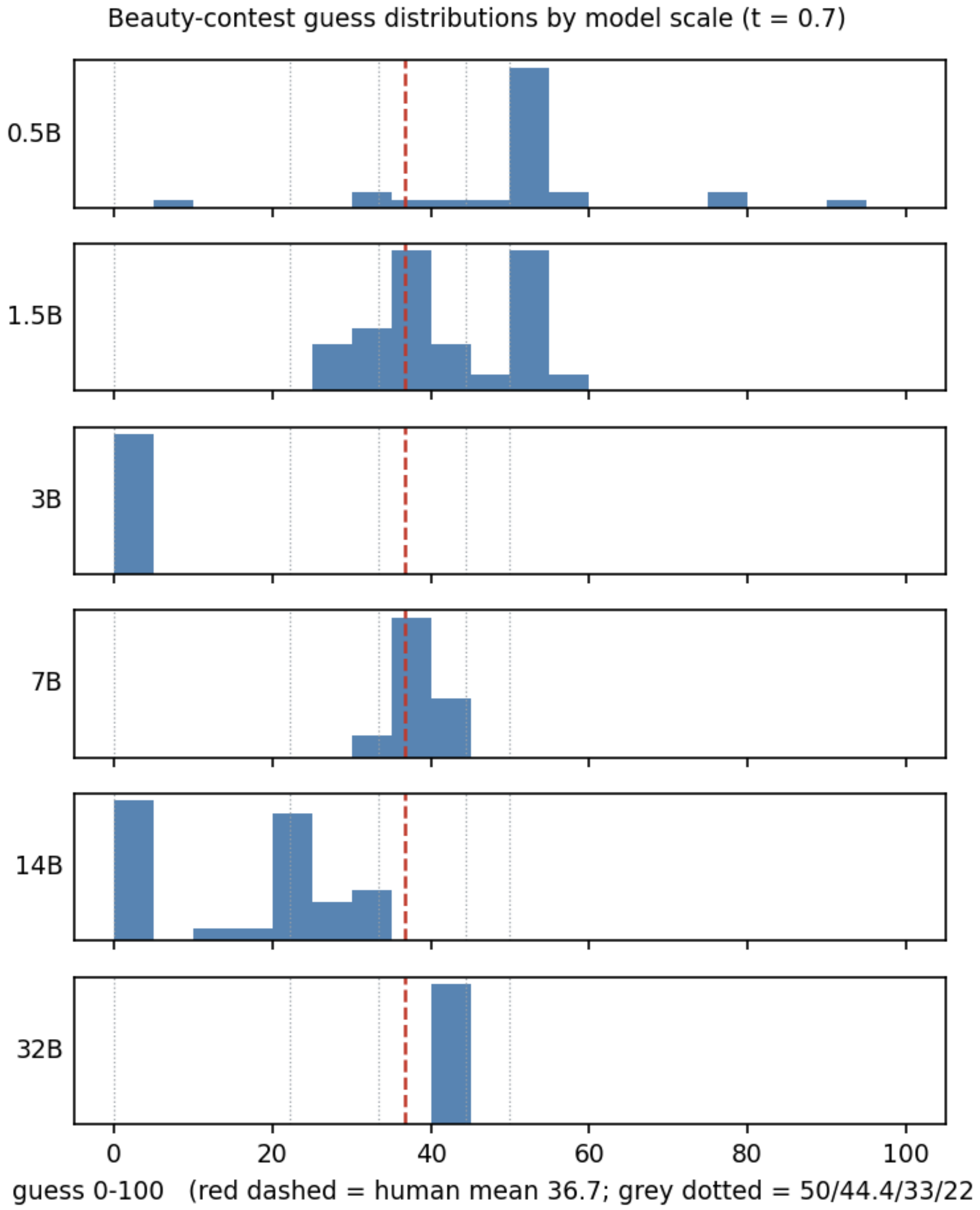

*Figure 1. Beauty-contest guess distributions by model scale (t = 0.7). Each model concentrates at one or two points (grey dotted lines at 50, 44.4, 33, 22, 0), where the human distribution spreads across them. The red dashed line marks the human mean of 36.7 [Nagel, 1995], which falls between the human low-level cluster near 22 to 33 and a smaller cluster near the level-0 anchor and above, a bimodality that no single model cell reproduces.*

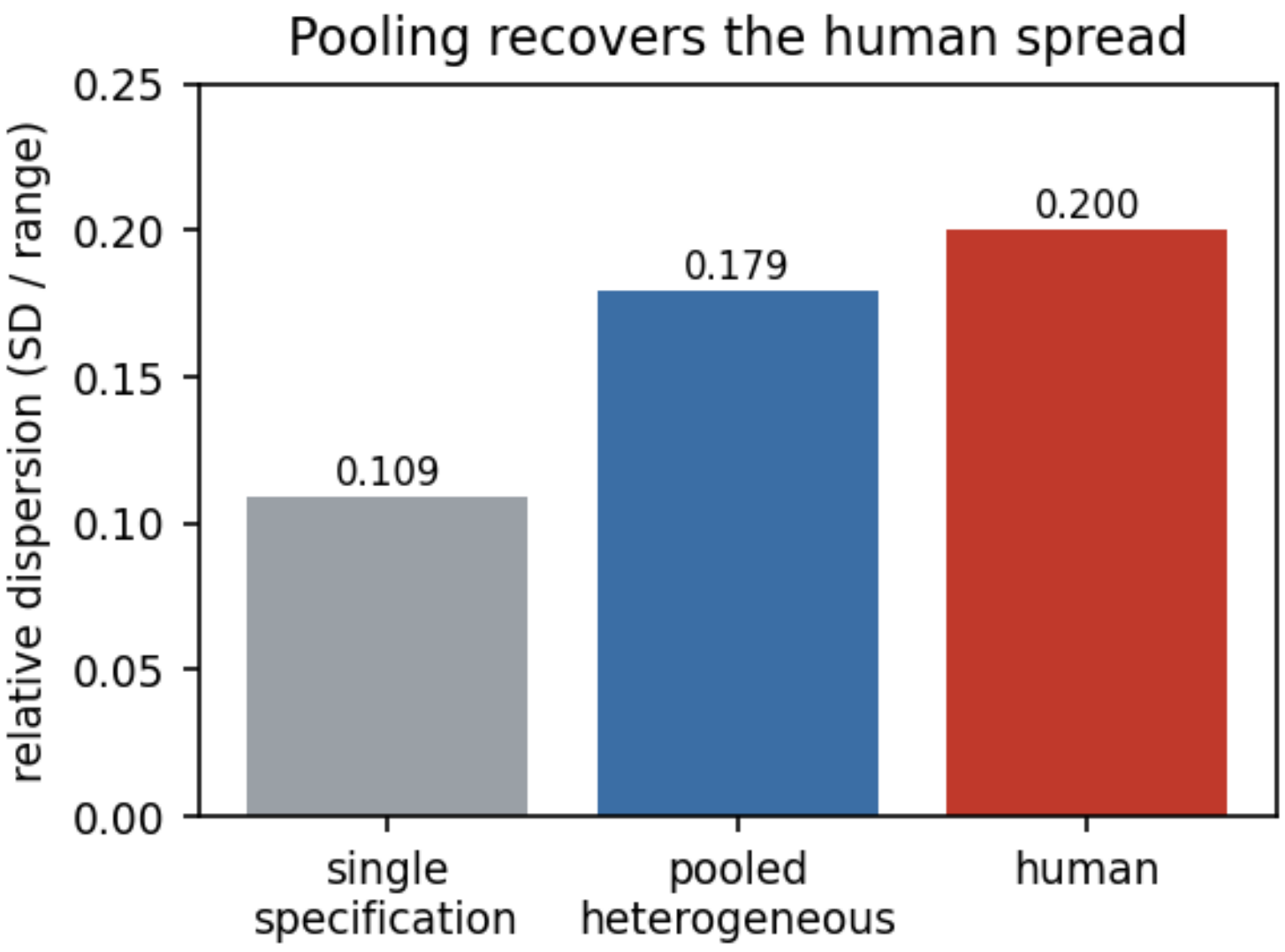

*Figure 2. Pooling heterogeneous specifications lifts the relative dispersion from 0.109 for a single specification to 0.179, approaching but not reaching the human value of 0.200.*

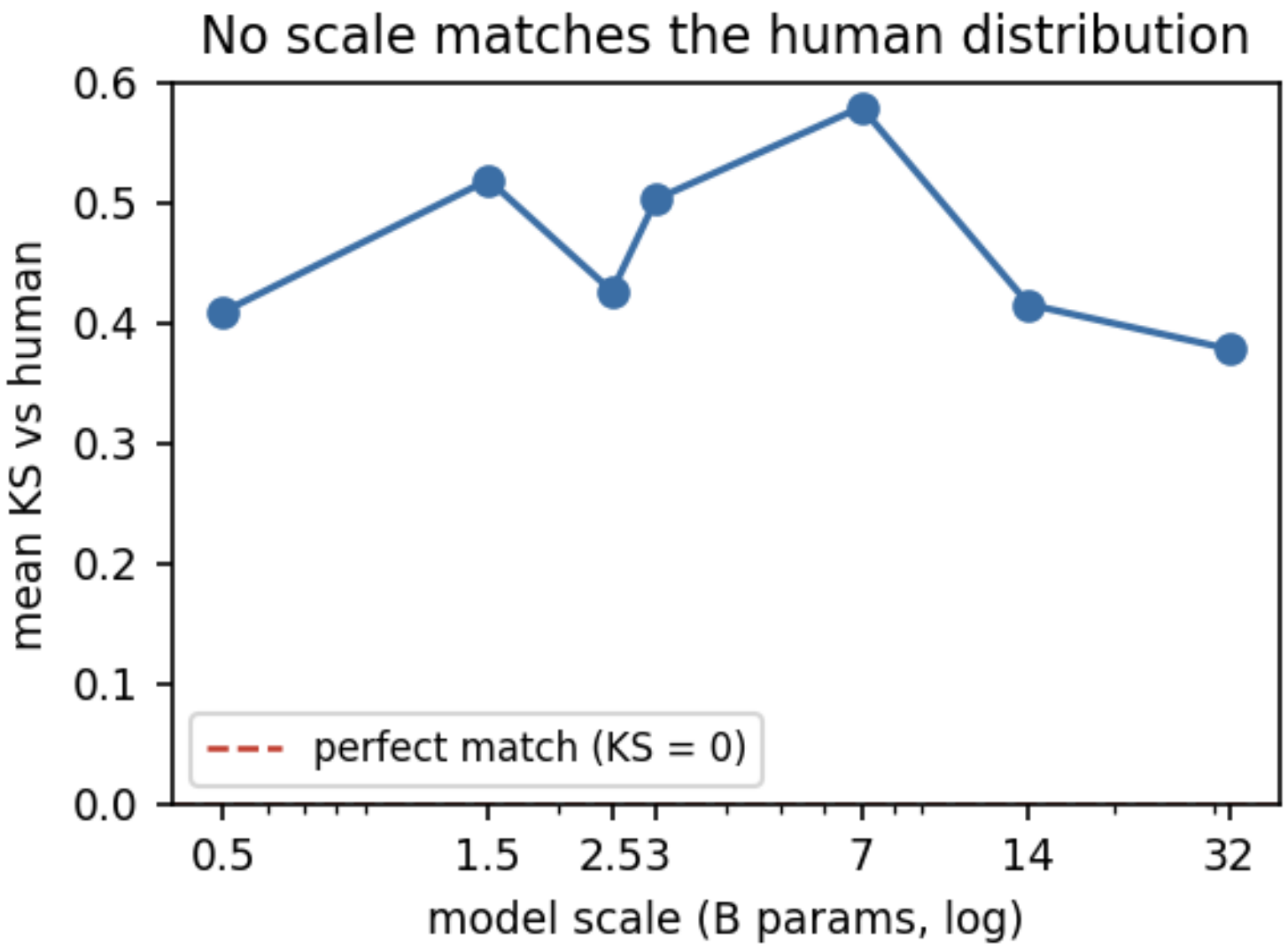

*Figure 3. Mean Kolmogorov-Smirnov distance between each model scale and the human distribution in the public goods game; no scale approaches the zero of a faithful match.*

In the next section, we show that none of the deployment settings restore the within-game dynamics that human play shows: the decline across rounds and the endgame unravelling. At the same time, the punishment-condition results, the inert sanctioning institution and the antisocial targeting of contributors, are reported in Appendix B. Any effects elicited by those mechanisms are attributable to the institution, rather than the level-k parameters mapped here.

# Results II: A Static, Category-Retrieved Level-k Player

Primary findings in this section suggest that for both games, the agent retrieves the reasoning level its training ties to the game's category. These findings are shown in the following subsections that test H1–H3: which level the agent plays, whether it plays that level statically, and whether it retrieves or computes it.

The level-k that synthetic agents play is varied with scale. In the beauty contest, the model ladder traces the level-k hierarchy [Ho et al., 1998] (see Figure 1). Aside from 3B and 32B, there is a monotone decrease in values as we move up the model ladder at t = 0.5, 0.7 and 1.0 (exact one-sided p = 0.042). For t = 0.2, rho is at -0.8 while for t = 1.3, rho = -0.4. Looking specifically at 3B, the ~0.91 value is due to a quantisation artifact (Q4). At q8 however, 3B's value goes to ~44.51 and at fp16, 39.16. 32B also exhibits a different pattern, hitting 44.44 at every temperature (SD ~ 0.0014), close to 3B's q8 value. Translating the values into level-k, a naive anchor at 50 will be a level-0, a level-1 will be (2/3) x 50 = 33.3, a level-2 at 22.2, and a level-3 at 14.8. Nagel [1995] placed the human mean at 36.7.

The public goods game’s analogous depth is the conditional-cooperation slope. The slope rises monotonically with scale at every temperature (1.5B ~0.12–0.23, 7B ~0.25–0.34, 32B ~0.47–0.70; see Table 2). The 32B model sits in the human band 0.5–0.7 [Fischbacher et al., 2001]. On the other hand, smaller models play a fixed level regardless of others. There is thus evidence from both paradigms that model parameter capability governs strategic depth. Across both games, H1 holds from the 0.5B–14B models, with the beauty contest ordering bounded before 32B, which instead fixes on a constant. For the public goods game, we find stronger evidence of the model capability pattern, with every scale in Table 2 following it.

| Model scale | t = 0.2 | t = 0.5 | t = 0.7 | t = 1.0 | t = 1.3 |
|---|---|---|---|---|---|
| 1.5B | 0.21 (0.07) | 0.23 (0.07) | 0.18 (0.07) | 0.12 (0.04) | 0.14 (0.05) |
| 7B | 0.31 (0.17) | 0.34 (0.18) | 0.27 (0.17) | 0.25 (0.16) | 0.25 (0.17) |
| 32B | 0.55 (0.15) | 0.70 (0.07) | 0.57 (0.09) | 0.49 (0.08) | 0.47 (0.08) |

*Table 2. Conditional-cooperation slope from the strategy method, by model scale and temperature, with the standard error of the slope in parentheses. Each slope is an OLS regression of the agent's best-response contribution based on the stated average contribution of other agents, elicited over an eleven-point grid spanning 0 to 20 (steps of 2), with twenty repetitions at each grid point. For comparison, the human conditional-cooperator band is 0.5 to 0.7 [Fischbacher et al., 2001].*

There is no evidence of dynamic level-k in synthetic agents. In the public goods game, no deployment setting combination was able to reproduce the human decline (-0.91 tokens/round). The steepest decline was -0.63 tokens/round, and the few cells that decline at all are a single small-model artefact (3B models under competitive framing). While some of the deployment settings do move the round trend as shown in the previous section, the most optimised combination settings still tend towards the flat-to-rising trend rather than the human decline in contributions across rounds. The slope gaps (see Table 3) also show the lack of human distributional faithfulness. This suggests that while a level-1 best-response function is present, the best response is not appropriately updated within rounds. As such, this confirms H2.

| Model scale | Cells | Mean KS | Mean Wasserstein | LLM decay slope | Slope gap (LLM minus human) |
|---|---|---|---|---|---|
| 0.5B | 61 | 0.412 | 4.75 | +0.243 | +1.153 |
| 1.5B | 61 | 0.486 | 5.79 | +0.129 | +1.039 |

| Model scale | Cells | Mean KS | Mean Wasserstein | LLM decay slope | Slope gap (LLM minus human) |
|---|---|---|---|---|---|
| 3B | 61 | 0.449 | 5.06 | +0.191 | +1.101 |
| 7B | 61 | 0.515 | 6.54 | +0.126 | +1.037 |
| 14B | 61 | 0.425 | 4.89 | +0.160 | +1.071 |
| 32B | 60 | 0.396 | 4.51 | +0.145 | +1.055 |

*Table 3. Distributional fidelity of the public goods game against the St. Gallen human data (condition N, n = 96 subjects per round), by model scale. The mean Kolmogorov-Smirnov statistic and mean Wasserstein distance are averaged over rounds 1 to 10, with lower values closer to the human distribution; the decay slope is in tokens per round, and the slope gap is the model slope minus the human slope of -0.91. Each row pools the N-condition cells at that scale across temperature, alignment, quantisation, and framing.*

The level is retrieved by category rather than computed. H3 is disentangled and measured through the horizon manipulation. Under an indefinite horizon, where round continuation is a stated probability, the 32B instruction-tuned model sustains more cooperation compared to a finite horizon, ranging from 1.8 to 3.3 tokens across temperatures. This finding is in line with the folk-theorem prediction. However, the horizon sensitivity in this comparison is more attributable to a single retrievable association rather than the local round position. Smaller models, and base models at every size, are flat across both horizon versions, defaulting to a single fixed response irrespective of the horizon. A known finite horizon should minimally imply a last-round defection by backward induction under classic Nash equilibrium. Real-world human play on the other hand softens this extreme to a partial drop in the final round. However the LLMs display neither Nash nor human strategic patterns. Even when the final round is made explicit, the 32B model still contributes its standing amount at every temperature but one. A single non-replicating elicitation is treated as noise in this study. Taken together, the model retrieves strategies based on the global horizon information, but does not run the within-game recursion, such as backward induction. Humans tend to under-recurse, yet synthetic agents do not recurse at all. The model therefore reproduces neither the Nash endgame defection nor the human partial drop. While it may seem that the finding here contradicts the last-round defection reported in frontier benchmarks [Affonso, 2026], those models in the benchmarks are above the local scale ceiling studied here, so the defection places backward induction as a potential capability beyond the scale range presented here, consistent with H1. Both findings, category sensitivity in the larger model and the absence of backward induction across the scales studied here, are what H3 predicts. H3 is thus confirmed: the capable model tracks the global horizon category, while no model across the scales studied here executes the local backward induction.

Findings in this section suggest that synthetic agents are static, category-retrieved level-k players. Scale sets which level-k the agent retrieves (H1), and the level is held static across rounds (H2). We also confirmed that strategies are retrieved by category rather than computed (H3). No deployment setting restores the within-game dynamics. Only by pooling heterogeneous specifications, can the human spread be recovered.

# Discussion

By studying a factorial of models from a single family, we acquire a clean map from controllable settings to behavioural coordinates, with each setting moving a different coordinate and quantisation moving none. Because the partition is clean, the map can be operationalised rather than merely noted. A researcher configuring an LLM subject pool can read off which setting shifts which behaviour, and which can be set aside.

Quantisation, while inert on average, still calls for some care. While its main effect is null on every coordinate in this study, there is a localised interaction with model scale, as noted by the distributional collapse at the 3B scale. However, that setting lacks systematic and continuous effects across deployment configurations. As long as a model is large enough, precision can thus be treated as a nuisance parameter for the population mean, provided it is held constant within a comparison.

Beyond deployment mechanisms, a behavioural game still requires a computational social science lens. After all, behavioural games are administered to identify key cognitive and decision-making processes. In the case of LLMs or synthetic agents, temperature is merely decision noise, and that setting does not represent human population heterogeneity. Researchers should not mistake trembling-hand phenomena around a mode [McKelvey & Palfrey, 1995] for the structure of human behavioural distributions. Synthetic agents by default do not assemble the mixture of reasoning types that gives human play its width [Camerer, Ho & Chong, 2004], where human play is better described by level-k cognitive and strategic reasoning. This is evident in the beauty contest simulations, where a model at a given scale anchors on a single level-k depth, a unimodal distribution, against the multimodal human distribution. This is not something that tuning a single deployment setting can simply emulate. The width within human data has to be constructed rather than sampled. Since the missing variation is between-subject rather than within-subject, the natural conclusion would be that varying the agents rather than the sampler would be the more appropriate approach.

To address agent heterogeneity beyond temperature, tuning the LLM's collective deployment settings in tandem, and employing appropriate contextual frames, somewhat supplies the between-agent spread. An assembled synthetic population is able to reach relative dispersion of 0.179, approaching the human value of 0.200 in this study, against the 0.109 a single specification reaches on its own. That way, both spread and structure can be simulated with greater faithfulness to human ground truth.

On model capability, the direction has already been established in existing literature. More capable models reason toward the game-theoretic solution while people deviate [Li & Shirado, 2025; Guzman Piedrahita et al., 2025]. Relying on increased model capability alone also leads simulated populations to concentrate more tightly than human ones, a recognised limitation of the approach [Mei et al., 2024]. The factorial in this study further affirms these findings. Specifically, scale lowers the settling level and flattens the upward drift without changing the opening. The equilibrium pull is thus a result of the trajectory rather than the result of the opening. The effect is also delineated from the prosocial level that alignment installs.

While we gleaned information from the deployment setting mechanisms, the response surface of the simulations alone does not explain the decision processes the players run. Horizon manipulation shows sensitivity to the global game category (indefinite versus finite) but not to local round position. At the same time, the strategy method shows that the level-1 best-response function is present yet static across the rounds. This implies that LLM agents retrieve a category-level response rather than computing from the local game progress. The characterisation of synthetic agents as static, category-retrieved level-k players is therefore established in this paper. What remains open is the selection of strategy or regularity that the model's training corpus installs in the category-to-level association. However, that is a question about model internals rather than behavioural characterisation, which the horizon and strategy-method findings settle.

The paper has also made deliberate choices to strengthen the trajectory results claim on a numerically bounded scale in behavioural games. Dispersion and opening behaviour are measured at the first round, before movements in subsequent rounds can compress them. Decline, in the case of the public goods game, is measured as a proportion of the opening rather than as a raw slope. This is done in order to refute the counterargument that a higher opening permits a steeper fall on a bounded scale. However, no cell in the factorial approaches the floor, and even the lowest settling level sits well inside the interval. The synthetic agents stop short of the boundary by choice and not by numerical censoring, and effects on decline from settings such as model temperature survive the normalised measure.

Taken together, the difference between human and LLM simulations can be analysed through the lens of bounded-rationality in strategic play. This is consistent with the proposed decomposition of human play into a strategic component, which the model reproduces, and a structured residual that the model does not [Teo, 2026]. This structured residual is not merely larger variance. It has structure, the mixture over reasoning depths and the conditional dependence on others realised through belief revision, that the design knobs move around but do not rebuild.

## Limitations

The model ladder is from a single dense family, so the scale findings hold the architecture and training corpus fixed. In order for the findings to be more generalisable, a cross-family slice is required to have more confidence that model capability is indeed scale rather than family.

This study also chose a battery of two games at opposite poles (a one-shot reasoning game versus a repeated social dilemma). This decision was made so that a common result across the two poles generalises in the cognitive analysis. However, two games may potentially confound horizon with valence and cannot separate their contributions. Comparison across multiple repeated games, for example, will further lend confidence to the horizon reading.

In this study, the human ground truth is derived from published distributions rather than fresh human samples. As such, this fixes the human targets but does not allow for finer-grain matching. Greater resolution data in human behavioural games allow for further nuances in the human modal proportions and comparisons to synthetic agents.

Lastly, the static, category-retrieved characterisation is supported by the horizon and strategy-method evidence in Results II. However, the characterisation describes the behaviour and its category-level control, not the model's internal computation. Architecture-general and richer-battery versions remain as future work.

In conclusion, the model studied in this paper is only faithful to human strategic play in a narrow sense. It retrieves a trained reasoning level that is associated with a game's category, fixes its position on the level-k ladder, and holds that level without the within-game computation that produces human strategic phenomena such as contribution decline in public goods games. The parameters of this static synthetic player are mapped by the controllable deployment settings. However, the human dispersion that behavioural game administrators are interested in is a mixture over reasoning levels rather than unimodal noise. Heterogeneous specifications across multiple simulated agents are thus required to achieve a more human-faithful mixture. While researchers remain motivated to engage in silicon-subject substitution, the agreement of means does not necessarily imply agreement of the strategic process that generated the human ground truth.

## Declarations

**Funding.** The author received no funding for this work.

**Conflict of interest.** On behalf of all authors, the corresponding author states that there is no conflict of interest.

**Ethics approval.** Not applicable. This study did not involve human participants or animals; all human data used are drawn from previously published studies.

**Consent to participate and publish.** Not applicable, as the study involved no human participants.

**Data and code availability.** The simulation harness, the analysis scripts, and the derived data panel that reproduce every statistic and figure reported here are available from the author on request.

**Author contributions.** The author confirms sole responsibility for the study conception and design, the simulations and analysis, and the preparation of the manuscript.

# References

Affonso, F. M. (2026). Large Language Models Converge on Competitive Rationality but Diverge on Cooperation across Providers and Generations. arXiv:2604.18596.

Aher, G., Arriaga, R. I., & Kalai, A. T. (2023). Using Large Language Models to Simulate Multiple Humans and Replicate Human Subject Studies. Proceedings of the 40th International Conference on Machine Learning (ICML), PMLR 202.

Akata, E., Schulz, L., Coda-Forno, J., Oh, S. J., Bethge, M., & Schulz, E. (2025). Playing Repeated Games with Large Language Models. Nature Human Behaviour.

Alekseenko, I., Dagaev, D., Paklina, S., & Parshakov, P. (2025). Strategizing with AI: Insights from a Beauty Contest Experiment. Journal of Economic Behavior & Organization. arXiv:2502.03158.

Argyle, L. P., Busby, E. C., Fulda, N., Gubler, J. R., Rytting, C., & Wingate, D. (2023). Out of One, Many: Using Language Models to Simulate Human Samples. Political Analysis, 31(3), 337-351.

Bosch-Domenech, A., Montalvo, J. G., Nagel, R., & Satorra, A. (2002). One, Two, (Three), Infinity, ...: Newspaper and Lab Beauty-Contest Experiments. American Economic Review, 92(5), 1687-1701.

Camerer, C. F., Ho, T.-H., & Chong, J.-K. (2004). A Cognitive Hierarchy Model of Games. Quarterly Journal of Economics, 119(3), 861-898.

Charness, G., & Rabin, M. (2002). Understanding Social Preferences with Simple Tests. Quarterly Journal of Economics, 117(3), 817-869.

Fehr, E., & Gachter, S. (2002). Altruistic Punishment in Humans. Nature, 415, 137-140.

Fischbacher, U., Gachter, S., & Fehr, E. (2001). Are People Conditionally Cooperative? Evidence from a Public Goods Experiment. Economics Letters, 71(3), 397-404.

Fischbacher, U., & Gachter, S. (2010). Social Preferences, Beliefs, and the Dynamics of Free Riding in Public Goods Experiments. American Economic Review, 100(1), 541-556.

Guzman Piedrahita, D., Yang, Y., Sachan, M., Ramponi, G., Scholkopf, B., & Jin, Z. (2025). Corrupted by Reasoning: Reasoning Language Models Become Free-Riders in Public Goods Games. arXiv:2506.23276.

Herrmann, B., Thoni, C., & Gachter, S. (2008). Antisocial Punishment Across Societies. Science, 319(5868), 1362-1367.

Ho, T.-H., Camerer, C., & Weigelt, K. (1998). Iterated Dominance and Iterated Best Response in Experimental p-Beauty Contests. American Economic Review, 88(4), 947-969.

Horton, J. J. (2023). Large Language Models as Simulated Economic Agents: What Can We Learn from Homo Silicus? NBER Working Paper 31122.

Huynh, T.-K., et al. (2025). Understanding LLM Agent Behaviours via Game Theory: Strategy Recognition, Biases and Multi-Agent Dynamics. arXiv:2512.07462.

Isaac, R. M., & Walker, J. M. (1988). Group Size Effects in Public Goods Provision: The Voluntary Contributions Mechanism. Quarterly Journal of Economics, 103(1), 179-199.

Jia, J., Yuan, Z., Pan, J., McNamara, P. E., & Chen, D. (2025). LLM Strategic Reasoning: Agentic Study through Behavioral Game Theory. arXiv:2502.20432.

Kahneman, D., Knetsch, J. L., & Thaler, R. H. (1991). Anomalies: The Endowment Effect, Loss Aversion, and Status Quo Bias. Journal of Economic Perspectives, 5(1), 193-206.

Li, Y., & Shirado, H. (2025). Spontaneous Giving and Calculated Greed in Language Models. EMNLP 2025. arXiv:2502.17720.

Liberman, V., Samuels, S. M., & Ross, L. (2004). The Name of the Game: Predictive Power of Reputations versus Situational Labels in Determining Prisoner's Dilemma Game Moves. Personality and Social Psychology Bulletin, 30(9), 1175-1185.

McKelvey, R. D., & Palfrey, T. R. (1995). Quantal Response Equilibria for Normal Form Games. Games and Economic Behavior, 10(1), 6-38.

Mei, Q., Xie, Y., Yuan, W., & Jackson, M. O. (2024). A Turing Test of Whether AI Chatbots Are Behaviorally Similar to Humans. Proceedings of the National Academy of Sciences, 121(9), e2313925121.

Nagel, R. (1995). Unraveling in Guessing Games: An Experimental Study. American Economic Review, 85(5), 1313-1326.

Teo, P. H. (2026). Divergent Minds, Convergent Baselines: A Bounded-Rationality Account of LLM-Human Strategic Behaviour. arXiv:2605.26437.

Zelmer, J. (2003). Linear Public Goods Experiments: A Meta-Analysis. Experimental Economics, 6(3), 299-310.

# Appendix A: Agent Prompts

*Generated verbatim from the simulation code (`sims/games/`). Every prompt an agent received is reproduced exactly; no human data ever appears in any prompt. Bracketed amounts are the model's own history, never experimental ground truth.*

## A. p-beauty contest

One-shot, p = 2/3, 18 players. Single prompt per agent.

`

You are one of 18 players in a guessing game.

Each player picks a number between 0 and 100 (decimals allowed).

The winner is the player whose number is closest to 0.6667 times the average of all 18 chosen numbers.

Think about what number gives you the best chance to win, then state it.

Respond with ONLY your chosen number on the last line, like:

ANSWER: <number>

`

## B. Public goods (VCM) — contribution decision

Instruction-tuned path. The mechanics block is identical across framings; only the intro and decision line carry the contextual valence.

### *Neutral framing (round 1, no history)*

`

You are taking part in a 4-person decision task over 10 rounds.

Each round you receive 20 tokens and choose an integer amount (0 to 20) to place in a shared pot; the rest stays in your private account.

Private-account tokens are kept as they are. Every token any of the 4 participants places in the shared pot is multiplied and split equally among all 4, so each pot token returns 0.4 to each of you.

The same 4 participants stay together for all 10 rounds. After everyone decides, you are shown how much each other participant placed in the pot.

You are participant P1.

Decide how many of your 20 tokens to place in the shared pot this round.

Respond with ONLY your amount on the last line, like:

ANSWER: <0-20>

`

### *Cooperative framing (round 1, no history)*

`

You are one of 4 teammates working together over 10 rounds. Building up the shared pot helps the whole group, and the more the team puts in together, the better everyone does.

Each round you receive 20 tokens and choose an integer amount (0 to 20) to place in a shared pot; the rest stays in your private account.

Private-account tokens are kept as they are. Every token any of the 4 participants places in the shared pot is multiplied and split equally among all 4, so each pot token returns 0.4 to each of you.

The same 4 participants stay together for all 10 rounds. After everyone decides, you are shown how much each other participant placed in the pot.

You are participant P1.

Decide how many of your 20 tokens to contribute to your group this round.

Respond with ONLY your amount on the last line, like:

ANSWER: <0-20>

`

### *Competitive framing (round 1, no history)*

`

You are one of 4 players competing over 10 rounds to earn the most tokens for yourself. Tokens you keep in your private account are guaranteed yours; tokens you place in the pot are shared with your rivals.

Each round you receive 20 tokens and choose an integer amount (0 to 20) to place in a shared pot; the rest stays in your private account.

Private-account tokens are kept as they are. Every token any of the 4 participants places in the shared pot is multiplied and split equally among all 4, so each pot token returns 0.4 to each of you.

The same 4 participants stay together for all 10 rounds. After everyone decides, you are shown how much each other participant placed in the pot.

You are participant P1.

Decide how many of your 20 tokens to keep for yourself versus place in the pot this round, aiming to maximise your own earnings.

Respond with ONLY your amount on the last line, like:

ANSWER: <0-20>

`

### *B.H neutral framing with history (mid-game)*

`

You are taking part in a 4-person decision task over 10 rounds.

Each round you receive 20 tokens and choose an integer amount (0 to 20) to place in a shared pot; the rest stays in your private account.

Private-account tokens are kept as they are. Every token any of the 4 participants places in the shared pot is multiplied and split equally among all 4, so each pot token returns 0.4 to each of you.

The same 4 participants stay together for all 10 rounds. After everyone decides, you are shown how much each other participant placed in the pot.

You are participant P1.

What has happened so far:

Round 1: you put 12 in the pot, the others put [10, 14, 8], you earned 25.6 tokens.

Round 2: you put 10 in the pot, the others put [9, 11, 6], you earned 24.4 tokens.

Decide how many of your 20 tokens to place in the shared pot this round.

Respond with ONLY your amount on the last line, like:

ANSWER: <0-20>

`

## C. Horizon arms (mechanism study; neutral framing, round 1)

Identical except the horizon clause. Tests retrieval (invariance) vs reasoning (tracks structure).

### *Finite horizon*

`

You are taking part in a 4-person decision task over 10 rounds.

Each round you receive 20 tokens and choose an integer amount (0 to 20) to place in a shared pot; the rest stays in your private account.

Private-account tokens are kept as they are. Every token any of the 4 participants places in the shared pot is multiplied and split equally among all 4, so each pot token returns 0.4 to each of you.

The same 4 participants stay together for all 10 rounds. After everyone decides, you are shown how much each other participant placed in the pot.

You are participant P1.

Decide how many of your 20 tokens to place in the shared pot this round.

Respond with ONLY your amount on the last line, like:

ANSWER: <0-20>

`

### *Indefinite horizon*

`

You are taking part in a 4-person decision task over a series of rounds with no fixed end announced in advance.

Each round you receive 20 tokens and choose an integer amount (0 to 20) to place in a shared pot; the rest stays in your private account.

Private-account tokens are kept as they are. Every token any of the 4 participants places in the shared pot is multiplied and split equally among all 4, so each pot token returns 0.4 to each of you.

The same 4 participants stay together each round. After each round the task continues for another round with probability 0.8 (80% per round) and otherwise ends; the total number of rounds is not fixed in advance. After everyone decides, you are shown how much each other participant placed in the pot.

You are participant P1.

Decide how many of your 20 tokens to place in the shared pot this round.

Respond with ONLY your amount on the last line, like:

ANSWER: <0-20>

`

### *Unframed horizon*

`

You are taking part in a 4-person decision task over a series of rounds.

Each round you receive 20 tokens and choose an integer amount (0 to 20) to place in a shared pot; the rest stays in your private account.

Private-account tokens are kept as they are. Every token any of the 4 participants places in the shared pot is multiplied and split equally among all 4, so each pot token returns 0.4 to each of you.

The same 4 participants stay together each round. After everyone decides, you are shown how much each other participant placed in the pot.

You are participant P1.

Decide how many of your 20 tokens to place in the shared pot this round.

Respond with ONLY your amount on the last line, like:

ANSWER: <0-20>

`

## D. Strategy-method conditional-cooperation elicitation

Single-shot; the stated others' contribution is varied across 0..20 to trace the best-response schedule.

`

You are taking part in a 4-person decision task over 10 rounds.

Each round you receive 20 tokens and choose an integer amount (0 to 20) to place in a shared pot; the rest stays in your private account.

Private-account tokens are kept as they are. Every token any of the 4 participants places in the shared pot is multiplied and split equally among all 4, so each pot token returns 0.4 to each of you.

The same 4 participants stay together for all 10 rounds. After everyone decides, you are shown how much each other participant placed in the pot.

You are participant P1. Suppose the other 3 participants each place 12 tokens in the pot this round.

Decide how many of your 20 tokens to place in the shared pot this round.

Respond with ONLY your amount on the last line, like:

ANSWER: <0-20>
`

## E. Belief elicitation

`

You are taking part in a 4-person decision task over 10 rounds.

Each round you receive 20 tokens and choose an integer amount (0 to 20) to place in a shared pot; the rest stays in your private account.

Private-account tokens are kept as they are. Every token any of the 4 participants places in the shared pot is multiplied and split equally among all 4, so each pot token returns 0.4 to each of you.

The same 4 participants stay together for all 10 rounds. After everyone decides, you are shown how much each other participant placed in the pot.

You are participant P1.

What has happened so far:

Round 1: you put 12 in the pot, the others put [10, 14, 8], you earned 25.6 tokens.

Round 2: you put 10 in the pot, the others put [9, 11, 6], you earned 24.4 tokens.

Predict what the other 3 participants will place in the pot next round, on average.

Respond with ONLY your prediction on the last line, like:

ANSWER: <0-20>
`

## F. Endgame salience (final-round note)

`

You are taking part in a 4-person decision task over 10 rounds.

Each round you receive 20 tokens and choose an integer amount (0 to 20) to place in a shared pot; the rest stays in your private account.

Private-account tokens are kept as they are. Every token any of the 4 participants places in the shared pot is multiplied and split equally among all 4, so each pot token returns 0.4 to each of you.

The same 4 participants stay together for all 10 rounds. After everyone decides, you are shown how much each other participant placed in the pot.

You are participant P1.

What has happened so far:

Round 1: you put 12 in the pot, the others put [10, 14, 8], you earned 25.6 tokens.

Round 2: you put 10 in the pot, the others put [9, 11, 6], you earned 24.4 tokens.

This is the final round.

Decide how many of your 20 tokens to place in the shared pot this round.

Respond with ONLY your amount on the last line, like:

ANSWER: <0-20>

`

## G. Punishment decision (P condition; four wordings, identical payoff)

### *'punishment' wording*

`

You are participant P1. This round you put 10 in the pot. The others put:

Member A: put 14 in the pot

Member B: put 8 in the pot

Member C: put 11 in the pot

You may now punish any of them. Assign 0 to 10 punishment points to each member. Each one you assign costs you 1 token and removes 3 tokens from that member. They will not know who punished them.

Respond with ONLY your assignment on the last line, like:

ANSWER: A=<0-10>, B=<0-10>, C=<0-10>

`

### *'deduction' wording*

`

You are participant P1. This round you put 10 in the pot. The others put:

Member A: put 14 in the pot

Member B: put 8 in the pot

Member C: put 11 in the pot

You may now deduct from any of them. Assign 0 to 10 deduction points to each member. Each one you assign costs you 1 token and removes 3 tokens from that member. They will not know who deducted from them.

Respond with ONLY your assignment on the last line, like:

ANSWER: A=<0-10>, B=<0-10>, C=<0-10>

`

### *'fee' wording*

`

You are participant P1. This round you put 10 in the pot. The others put:

Member A: put 14 in the pot

Member B: put 8 in the pot

Member C: put 11 in the pot

You may now charge any of them. Assign 0 to 10 fees to each member. Each one you assign costs you 1 token and removes 3 tokens from that member. They will not know who charged them.

Respond with ONLY your assignment on the last line, like:

ANSWER: A=<0-10>, B=<0-10>, C=<0-10>

`

### *'points' wording*

`

You are participant P1. This round you put 10 in the pot. The others put:

Member A: put 14 in the pot

Member B: put 8 in the pot

Member C: put 11 in the pot

You may now assign points to any of them. Assign 0 to 10 points to each member. Each one you assign costs you 1 token and removes 3 tokens from that member. They will not know who assigned points to them.

Respond with ONLY your assignment on the last line, like:

ANSWER: A=<0-10>, B=<0-10>, C=<0-10>

`

## H. Base-model completion path (structural axis)

Base models complete text rather than follow instructions, so the prompt ends mid-sentence before the number. PRIMES injects a social cue to measure copying.

### *H.none neutral, no prime (round 1)*

`

You are taking part in a 4-person decision task over 10 rounds.

Each round you receive 20 tokens and choose an integer amount (0 to 20) to place in a shared pot; the rest stays in your private account.

Private-account tokens are kept as they are. Every token any of the 4 participants places in the shared pot is multiplied and split equally among all 4, so each pot token returns 0.4 to each of you.

The same 4 participants stay together for all 10 rounds. After everyone decides, you are shown how much each other participant placed in the pot.

Round 1: participant P1 places

`

### *Neutral, 'coop' prime (round 1)*

`

You are taking part in a 4-person decision task over 10 rounds.

Each round you receive 20 tokens and choose an integer amount (0 to 20) to place in a shared pot; the rest stays in your private account.

Private-account tokens are kept as they are. Every token any of the 4 participants places in the shared pot is multiplied and split equally among all 4, so each pot token returns 0.4 to each of you.

The same 4 participants stay together for all 10 rounds. After everyone decides, you are shown how much each other participant placed in the pot.

In earlier sessions, participants usually placed almost all of their tokens (around 18 to 20) in the pot.

Round 1: participant P1 places

`

### *Neutral, 'defect' prime (round 1)*

`

You are taking part in a 4-person decision task over 10 rounds.

Each round you receive 20 tokens and choose an integer amount (0 to 20) to place in a shared pot; the rest stays in your private account.

Private-account tokens are kept as they are. Every token any of the 4 participants places in the shared pot is multiplied and split equally among all 4, so each pot token returns 0.4 to each of you.

The same 4 participants stay together for all 10 rounds. After everyone decides, you are shown how much each other participant placed in the pot.

In earlier sessions, participants usually placed almost none of their tokens (around 0 to 2) in the pot.

Round 1: participant P1 places

`

# Appendix B: The Punishment Condition

The original human ground-truth study included both a punishment condition and a non-punishment control. This punishment condition is reported here rather than in the body because its findings are a property of the sanctioning institution and its mechanisms, as opposed to the level-k parameters elicited from the main results. The punishment condition adds a second stage to each round of the public goods game: a participant, having seen the others' contributions, may assign costly sanctions to them. The punishment mechanism promotes cooperation in human subjects, increasing final-round contribution by roughly twelve tokens [Fehr & Gachter, 2002; Herrmann et al., 2008]. In the LLM simulations, the sanctioning institution does not follow the human ground truth. The models punish the wrong target and the targeting is neither retaliatory nor a result of the sanction's wording.

While in the human study, the final-round contribution rises by about twelve tokens once punishment is available, the swing in synthetic agents runs the other way, between -0.25 and -2.32 tokens across the ladder. Punishment reduces contribution slightly rather than sustaining it. The sanctioning institution that carries the paradigm in humans therefore produces no cooperative gain in the agents, at any scale tested.

| Model scale | Final round, no punishment | Final round, with punishment | Swing (P minus N) |
|---|---|---|---|
| 0.5B | 9.78 | 9.53 | -0.25 |
| 1.5B | 12.99 | 10.79 | -2.20 |
| 3B | 11.45 | 9.12 | -2.32 |
| 7B | 15.56 | 13.59 | -1.97 |
| 14B | 12.47 | 10.29 | -2.18 |
| 32B | 8.01 | 7.50 | -0.51 |

*Table B1. Final-round mean contribution without and with the punishment stage, by model scale (instruction-tuned, neutral framing). The human swing is about +12 (3.0 to 15.2) [Fehr & Gachter, 2002; Herrmann et al., 2008].*

By regressing punishment received on the target agent’s contribution, the slope is positive at intermediate scale (0.62 at 1.5B, 0.54 at 7B, 0.41 at 14B). Higher contributors therefore invite more punishment. This is similar to the antisocial pattern Herrmann et al. [2008] found to vary across human societies, which in turn neutralises the disciplinary value of sanctions.

However, this antisocial pattern is neither scale-monotonic nor uniform. The effect of contributions on punishment received is near zero at 3B, and it reverses at 32B (slope -0.25). The 32B model directs punishment at low contributors instead, which leans towards the prosocial direction. It is thus difficult to conclude the effect of model scale on punishment targeting, which is why it is presented here as a secondary observation.

| Model scale | Targeting slope, all rounds | Targeting slope, round 1 | Retaliation correlation |
|---|---|---|---|
| 0.5B | 0.07 | 0.13 | 0.02 |
| 1.5B | 0.62 | 1.41 | 0.02 |
| 3B | -0.01 | 0.01 | -0.03 |
| 7B | 0.54 | -0.54 | -0.13 |
| 14B | 0.41 | 1.04 | 0.12 |

| Model scale | Targeting slope, all rounds | Targeting slope, round 1 | Retaliation correlation |
|---|---|---|---|
| 32B | -0.25 | -0.30 | 0.04 |

*Table B2. Targeting slope is the regression of punishment received on the target's own contribution, so a positive value is antisocial targeting of high contributors. Retaliation correlation is the correlation between punishment received in one round and punishment sent in the next.*

Human antisocial punishment can be understood as retaliation. Free riders who are punished strike back at the cooperators who punished them. In the agents, the correlation between punishment received in one round and punishment dealt in the next is near zero at every scale, between -0.13 and +0.12, which suggests a lack of retaliatory computation between rounds. In models where the antisocial pattern is strongest on aggregate, the pattern is already present in round 1, before any sanctioning has occurred. Looking at round-1 plays alone, the targeting slope exceeds 1.0. Synthetic agents that punish antisocially do so reflexively, targeting the most conspicuous contributor right at cold-start, independent of historical information.

We also held the payoff fixed and varied only the label of the sanction (through wording such as “deduction”, “fee”, “points”, and “punishment”). This is done to disentangle the semantic effect of the word from the economic effect. However, controlling for this still produces no consistent ordering by valence. Using "punishment" as a loaded word does not depress contribution relative to the neutral "deduction", and in fact draws the highest contribution of the four at some model scales. The depression of cooperation under the sanction stage is therefore attributable to the economic cost of the sanction rather than the connotations of the word that describes the sanction mechanism.

| Model scale | Deduction | Fee | Points | Punishment |
|---|---|---|---|---|
| 1.5B | 10.79 | 9.78 | 9.34 | 11.08 |
| 7B | 13.59 | 16.03 | 13.07 | 12.57 |
| 14B | 10.29 | 7.73 | 5.92 | 10.70 |
| 32B | 7.50 | 5.49 | 5.87 | 7.98 |

*Table B3. Final-round mean contribution under four labels for the identical costly sanction. No ordering by emotional valence is consistent across scale.*

Two caveats bound the punishment results. First, the antisocial-targeting slope is unstable across the study, changing direction across scale and moving substantially across wordings. Therefore, it is reported as a secondary observation that warrants replication, rather than a settled conclusion.

The second caveat concerns the simulation parsing. The agent's punishment decision comes back as free text, which a parser converts into a number. Sometimes the parser is unable to extract a number. In these cases the parser conflates an unparsed response with an assignment of zero. As such, the findings on the absolute level of punishment can be treated as a lower bound. The zeroes are spurious. Since the targeting slope and retaliation correlation are relative measures, the spurious zeroes from parse failures do not affect those measures much.